\documentclass[12pt,preprint,psfig]{aastex}

\usepackage{emulateapj5}

\def\ltsima{$\; \buildrel < \over \sim \;$}

\def\lsim{\lower.5ex\hbox{\ltsima}}
\def\gtsima{$\; \buildrel > \over \sim \;$}
\def\gsim{\lower.5ex\hbox{\gtsima}}

\begin{document}
\title{Cosmic Variance In the Transparency of the
Intergalactic Medium After Reionization}

\author{J. Stuart B. Wyithe\altaffilmark{1} and Abraham
Loeb\altaffilmark{2}}

\email{swyithe@physics.unimelb.edu.au; aloeb@cfa.harvard.edu}

\altaffiltext{1}{University of Melbourne, Parkville, Victoria, Australia}

\altaffiltext{2}{Harvard-Smithsonian Center for Astrophysics, 60 Garden
St., Cambridge, MA 02138}

\begin{abstract}
Following the completion of cosmic reionization, the mean-free-path of
ionizing photons was set by a population of Ly-limit absorbers. As the
mean-free-path steadily grew, the intensity of the ionizing background
also grew, thus lowering the residual neutral fraction of hydrogen in
ionization equilibrium throughout the diffuse intergalactic medium
(IGM).  Ly$\alpha$ photons provide a sensitive probe for tracing the
distribution of this residual hydrogen at the end of
reionization. Here we calculate the cosmic variance among different
lines-of-sight in the distribution of the mean Ly$\alpha$ optical
depths. We find fractional variations in the effective
post-reionization optical depth that are of order unity on a scale of
$\sim100$ co-moving Mpc, in agreement with observations towards
high-redshift quasars. Significant contributions to these variations
are provided by the cosmic variance in the density contrast on the
scale of the mean-free-path for ionizing photons, and by fluctuations
in the ionizing background induced by delayed or enhanced structure
formation. Cosmic variance results in a highly asymmetric distribution
of transmission through the IGM, with fractional fluctuations in
Ly$\alpha$ transmission that are larger than in Ly$\beta$
transmission.

\end{abstract}


\section{Introduction}

It is thought that the appearance of the first galaxies and quasars
reionized the cosmic hydrogen in the intergalactic medium (IGM) that
was left neutral following the recombination era at $z\sim10^3$.  The
process started with the emergence of ionized regions around galaxies
and then groups of galaxies. These \ion{H}{2} regions eventually
overlapped, leaving the IGM highly ionized in low-density regions
surrounding pockets of residual neutral hydrogen in high density
regions. The process of reionization was inhomogeneous (Furlanetto et
al.~2004), and the overlap should have occurred at different times in
different regions (Wyithe \& Loeb~2004b). However the identification
of a Gunn-Peterson trough in the spectra of the most distant quasars
at redshifts of $z\sim 6.3$--$6.4$ (Fan et al.~2004) hints that the
reionization of cosmic hydrogen was completed near $z\sim6$, about a
billion years after the big bang (White et al. 2003; Wyithe \&
Loeb~2004a; Mesinger \& Haiman~2004).

Prior to the overlap epoch, the mean-free-path of ionizing photons was
limited by the size of individual \ion{H}{2} regions. Following
overlap, regions of IGM were exposed to ionizing radiation from
additional sources. This increase in the ionizing intensity ionized
hydrogen in denser regions, and resulted in an increased
mean-free-path which in turn led to an even higher ionizing intensity. As
the ionizing intensity in the post-overlap IGM rose, the optical depth
to Ly$\alpha$ absorption decreased (e.g. Fan et al.~2002).
Observations of the absorption spectra of high redshift quasars (Fan
et al. 2004) and potentially gamma-ray burst afterglows (Barkana \&
Loeb 2004; Chen et al. 2005) provide clues to the evolution of the
ionizing radiation field and of the evolution of the ionizing photon
mean-free-path.

The post-overlap evolution in the IGM is expected to be
inhomogeneous. Indeed values of optical depth to Ly$\alpha$ absorption
measured along different lines of sight at $5\la z\la6$ show
fluctuations that are of order unity (Becker et al.~2001; White et
al.~2003; Djorgovski~2004). The Ly$\alpha$ optical depth therefore
varies from place to place at any slice of cosmic time.  In addition
to small scale variations in Ly$\alpha$ optical depth generated by
structures like filaments and mini-halos (the so-called Ly$\alpha$
forest), there would be large scale inhomogeneities in the evolution
of the ionization state of the IGM. Fluctuations in the transmission
averaged on a large scale are sensitive to both small scale structure
along the line of sight (Lidz et al. 2005), and to cosmic variance on
larger scales. Here we concentrate on the contribution from cosmic
variance in the smoothed density field. However the presence of
high-column density absorbers and small scale structure in the density
field will add additional fluctuations to those described in this
paper.  Variations in the average transmission would be present on the
scale of the mean-free-path for ionizing photons ($\sim 10$--100
co-moving Mpc).  Indeed cosmic variance on scales of 10s of Mpc can be
significant and has not described by contemporary numerical
simulations of the reionization process (Barkana \& Loeb~2004). To
model cosmic variance properly with numerical simulations requires
resolving low-mass galaxies as well as solving the radiative transfer
equations inside a very large box (hundreds of co-moving Mpc) with a
high dynamic range of scales. Progress is now being made toward this
goal (Kohler, Gnedin, Miralda-Escud{\'e} \& Shaver 2005; Kohler,
Gnedin \& Hamilton~2005; Iliev et al.~2005).

In this paper we examine the variations in the optical depth averaged
over a redshift bin of sufficient size to smooth out the small-scale
inhomogeneities of the Ly$\alpha$ forest. Our model for the variation
in optical depth includes several contributions: {\it (i)} variations
in the density-contrast on the scale of the mean-free-path for
ionizing photons, {\it (ii)} variations in the ionizing background on
the same scale due to enhanced or delayed structure-formation, {\it
(iii)} variations in the ionizing background due to finite light
travel time across the clustering length of ionizing sources, {\it
(iv)} variations in the ionizing background due to Poisson
fluctuations in the number of ionizing sources, and {\it (v)}
evolution of the density of Ly-limit absorbers. Throughout the paper
we adopt the set of cosmological parameters determined by the {\em
Wilkinson Microwave Anisotropy Probe} (WMAP, Spergel et al. 2003),
namely mass density parameters of $\Omega_{m}=0.27$ in matter,
$\Omega_{b}=0.044$ in baryons, $\Omega_\Lambda=0.73$ in a cosmological
constant, and a Hubble constant of $H_0=71~{\rm
km\,s^{-1}\,Mpc^{-1}}$.

\section{Fluctuations in ionizing intensity and Ly$\alpha$ optical depth}

The background intensity of ionizing radiation $J_{\nu}$ is proportional to
the co-moving luminosity density due to ionizing sources ($n_{\rm src}$),
the density contrast ($\Delta$), the bias of sources in an overdense region
($B_{\rm src}$), the co-moving mean-free-path for ionizing photons ($R_{\rm
mfp}$), and a factor ($F_{\rm Q}$) describing the fractional contribution
of quasars to the ionizing back-ground relative to stars
\begin{equation}
J_{\nu} \propto n_{\rm src}(1+z)^3 \Delta B_{\rm src} \frac{R_{\rm mfp}}{1+z} (1+F_{\rm Q}).
\end{equation}
Note that we assume a universal spectrum for the contribution to
$J_\nu$ arising from a fixed redshift because we are averaging over a
large number of sources. However the contribution to the ionizing
background at a fixed point in the IGM from redshifted photons will
result in a value for $J_\nu$ that depends on the shape of this
universal spectrum, as well as the mean-free-path. On the other hand
this dependence on contribution from redshifted photons is the same in
all regions of size $R_{\rm mfp}$. Therefore since in this paper our
interest lies in relative fluctuations of the ionizing back-ground
among random regions of fixed size centered at the same redshift, we
assume the exact shape of the spectrum to be unimportant for our
calculation.

The mean-free-path ($R_{\rm mfp}$) is in turn related to the effective
co-moving density of Ly-limit absorbers ($\Delta\times n_{\rm abs}$) as well
as their bias ($B_{\rm abs}$)
\begin{equation}
\frac{R_{\rm mfp}}{(1+z)} \propto \left[n_{\rm abs}(1+z)^3 \Delta\times
B_{\rm abs}\right]^{-1}.
\end{equation}
We can also relate the average co-moving density of sources to the
collapsed fraction of baryons ($F_{\rm col}$) in dark-matter halos above
some minimum mass (that reflects the threshold for cooling and
fragmentation of the gas into stars or accreting black holes).  We
assume that ionizing photons were generated by starbursts, and
therefore expect that the luminosity density of ionizing photons was
proportional to the star-formation rate. We further assume that the
starbursts formed primarily out of newly collapsing IGM. As a result
the star-formation rate is assumed to be proportional to
the derivative of the collapsed fraction $n_{\rm src}
\propto (dF_{\rm col}/dt)$, so that
\begin{equation}
\label{J21}
J_{\nu} \propto \frac{(dF_{\rm col}/dt)}{n_{\rm abs}}\frac{B_{\rm src}}{B_{\rm abs}}(1+F_{\rm Q}).
\end{equation}
If both sources and absorbers reside in halos of the same mass, then
$B_{\rm abs} = B_{\rm src}$. This may be the case once the gas in
mini-halos below the cooling and fragmentation threshold, has been
photo-evaporated. However at high redshifts just after the overlap of
\ion{H}{2} regions, the mini-halos might provide a significant contribution
to the population of Ly-limit absorbers (Barkana \& Loeb 2002; Iliev,
Shapiro, \& Raga 2005).  In this case $B_{\rm src}>B_{\rm abs}$.
These bias factors are expected to evolve with time; $B_{\rm src}$ may
have a slow redshift dependence through a changing minimum mass for galaxy
formation while $B_{\rm abs}$ may have a value that changes with
redshift if the typical absorber mass changes with time (for example if
mini-halos are being evaporated) and so $B_{\rm abs}$ could be a
function of the evolving $J_{\nu}$.

Suppose that $n_{\rm abs}$ is some general function of $F_{\rm col}$ and
$J_{\nu}$.  Ly-limit systems are expected to shrink in physical
cross-section as their lower-density envelopes are ionized by a growing
ionizing background. The effective density of absorbers should therefore be
a function of $J_{\nu}$ in addition to $F_{\rm col}$.  Moreover since
$F_{\rm col}$ and $(dF_{\rm col}/dt)$ are both monotonic there is a unique
one-to-one correspondence between the values of $F_{\rm col}$ and $(dF_{\rm
col}/dt)$ as a function of time. Equation~(\ref{J21}) is therefore an
implicit equation for $J_{\nu}$ as a function of $F_{\rm col}$. The above
considerations lead to two interesting conclusions. First, we find that
$J_{\nu}$ should be independent of density contrast at a fixed value of
collapsed fraction; and second, the average ionizing background $J_{\nu}$
within a region of radius equal to $R_{\rm mfp}$ should be a function only
of the value for $F_{\rm col}$. With these points in mind we can estimate
the scatter in the redshift at which a threshold value of $J_{\nu}$ is
reached, and from it find the corresponding scatter in the value of
$J_{\nu}$ at fixed redshift.

\begin{figure*}[t]
\epsscale{2.}  \plotone{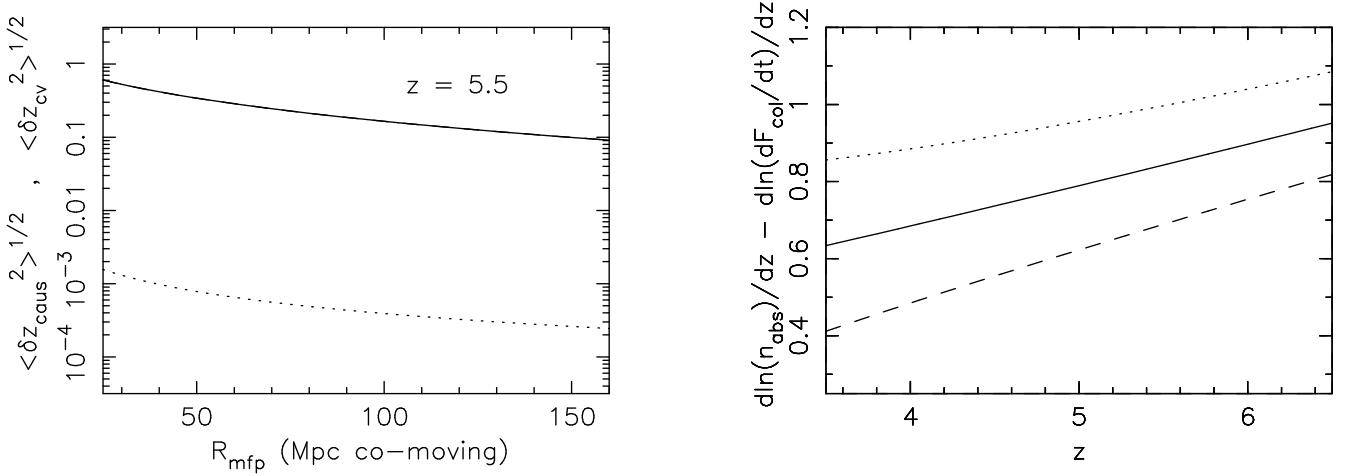}
\caption{\label{fig1} {\em Left panel:} Contribution to scatter in redshift
from cosmic variance $\langle\delta z_{\rm cv}^2\rangle^\frac{1}{2}$
(solid line) and finite light travel time $\langle\delta z_{\rm
caus}^2\rangle^\frac{1}{2}$ (dotted line) as a function of the
mean-free-path for ionizing photons in co-moving Mpc at a redshift of
5.5. {\em Right panel:} The value of $d\ln{n_{\rm abs}}/dz - d\ln{(dF_{\rm
col}/dt)}/dz$ as a function of redshift. The dashed, solid and dotted
lines correspond to $\gamma=2$, 3, and 4 respectively. In calculating
$F_{\rm col}$ we have assumed a minimum virial temperature for galaxies of
$2\times10^5$K. }
\end{figure*}

\subsection{Fluctuations due to cosmic variance}

Large-scale inhomogeneity in the cosmic density field leads to
structure-formation that is enhanced in over-dense regions and delayed
in under-dense regions.  The resulting cosmic variance in the redshift at which
a critical value of $F_{\rm col}$ is reached within regions of size
$R$ may be calculated as
\begin{equation}
\label{delta}
\langle\delta z_{\rm cv}^2\rangle^{\frac{1}{2}} =
\frac{\sigma_R}{\delta_c}(1+z),
\end{equation}
where $\sigma_{R}$ is the r.m.s. amplitude of the linear density field
smoothed over spheres of radius $R$, and $\delta_c$ is a critical overdensity
for collapse ($\propto (1+z)$ at high redshift; Barkana \& Loeb~2004; Bond
et al.~1991). Since the ionizing background within a given region depends
on the value of the collapsed fraction, we can relate $\langle\delta z_{\rm
cv}^2\rangle^{\frac{1}{2}}$ to the cosmic variance in redshift where a
critical value of $J_{\nu}$ is obtained.  The value of $J_{\nu}$ is
independent of $\Delta$. We therefore use the mean-free-path averaged over
all over-densities as the appropriate length scale over which the cosmic
variance is calculated, i.e. $R=R_{\rm mfp}/2$.

The left panel of Figure~\ref{fig1} shows the scatter in the redshift at
which a critical value of $J_{\nu}$ is obtained due to cosmic variance,
$\langle\delta z_{\rm cv}^2\rangle$ (solid lines).  We would like to
convert this scatter in redshift to a scatter in the optical-depth to
Ly$\alpha$ absorption at a given redshift.  At a fixed redshift $z$, the
local value of Ly$\alpha$ optical depth ($\tau_{\rm l}$) scales as
\begin{equation}
\tau_{\rm l} \propto \rho_{\rm HI}^2 J_{\nu}^{-1},
\end{equation}
where $\rho_{\rm HI}$ is the density of neutral hydrogen (Barkana \&
Loeb~2004). The effective optical depth ($\tau$) is measured over a
path-length through the IGM, and may be computed from $\tau=-\log{F}$
where $F=\langle e^{-\tau_{\rm l}}\rangle$ is the transmission
averaged over the path-length. While the ionizing background will be
sensitive to cosmic variance in density contrast calculated in
3-dimensional volumes ($\Delta$) on the scale $R_{\rm mfp}$, the
absorption spectra of quasars probe only lines-of-sight through these
volumes. The effective optical depth averaged over a length scale
$R_{\rm mfp}$ therefore depends on the square of the density averaged
along the line-of-sight (Lidz, Oh \& Furlanetto~2005). Here we neglect
density fluctuations on small scales and approximate the average of
the density contrast squared using the smoothed density contrast
($\Delta_z$) calculated in a thin cylinder at redshift $z$. This
procedure underestimates the fluctuations in transmission (Lidz et
al.~2005). Note that although $\tau$ varies with $z$ due to the growth
factor of density perturbations, we focus here on the scatter in
$\tau$ at a fixed $z$.  The fractional change in the value of the
optical depth relative to the average, given a density contrast
($\Delta_z$), and a delay ($\delta z$) in the redshift where the
critical collapse fraction is reached is
\begin{eqnarray}
\nonumber
\delta_\tau\equiv\frac{\tau-\tau_{\rm av}}{\tau_{\rm av}} &=& \Delta_z^2 \frac{J_{\nu}(z_{\rm av})}{J_{\nu}(z_{\rm av})+\frac{dJ_{\nu}}{dz}\delta z}-1\\
 &=& \Delta_z^2 \left(1+\frac{d\ln J_{\nu}}{dz}\delta z\right)^{-1}-1
\end{eqnarray}
where $\tau_{\rm av}$ is the average value of optical depth at $z_{\rm
av}$.  But $J_{\nu} \propto (dF_{\rm col}/dt)\, n_{\rm
abs}^{-1}(B_{\rm src}/B_{\rm abs})(1+F_{\rm Q})$
[equation~\ref{J21}]. Within this expression we can only compute the
evolution of collapsed fraction, and therefore assume an evolution
specified by a parameter $\gamma$ and of the form $n_{\rm abs}(B_{\rm
abs}/B_{\rm src})(1+F_{\rm Q}) \propto (1+z)^\gamma$ for the remaining
quantities. Since we are considering a fractional change in redshift
that is much smaller than unity a power-law approximation is an
appropriate choice. With this addition we get
\begin{equation}
\label{tauav}
\nonumber
\delta_\tau = \Delta_z^2 \left(1+\left(\frac{d\ln (dF_{\rm col}/dt)}{dz}-\frac{\gamma}{(1+z)}\right) \delta z\right)^{-1}-1.
\end{equation}
We note that equation~(\ref{tauav}) is a first order expression for
$\delta_\tau$. This choice may be justified by the fact that
$\left[\frac{d\ln (dF_{\rm col}/dt)}{dz}-\frac{\gamma}{(1+z)}\right]$ has
an evolution which is nearly linear with redshift (see figure~\ref{fig1}).

At this point we specify the value of $\gamma$. One can observe the
average number of Ly-limit systems per unit co-moving length along the
line-of-sight $(R_{\rm LL})$. Given the observed distribution of column
densities for Ly-limit systems, this distance can be related to the
mean free path for Ly-limit photons $R_{\rm mfp}=R_{\rm LL}/\sqrt{\pi}$
(Miralda-Escude~2003). Storrie-Lombardi et al. (1994) measured the
abundance of Ly-limit systems per unit redshift at $z=4$, $dN_{\rm
LL}/dz\sim3.3$, leading to a co-moving  spacing of $R_{\rm
LL}=[(1+z)c(dt/dz)/3.3]$ 
for Ly-limit systems along the line-of-sight. The resulting mean-free-path
for Ly-limit photons is $R_{\rm mfp}\sim120$ co-moving Mpc at $z=4$. This
value provides an upper limit on the mean-free-path for Ly-limit photons at
higher redshifts when the cosmic gas is denser and less ionized. The
evolution of the number of Ly-limit systems per redshift is $dN_{\rm
LL}/dz\propto(1+z)^\beta$ with $\beta\sim1.5$ (Storrie-Lombardi et
al. 1994). The evolution in the number of lines per unit co-moving length
$x$ is therefore
\begin{equation}
\frac{dN_{\rm LL}}{dx} \propto (1+z)^{-1}\frac{dN_{\rm LL}}{dz}\left(\frac{cdt}{dz}\right)^{-1}\propto(1+z)^{\beta+\frac{3}{2}},
\end{equation}
where $c$ is the speed of light and we have neglected the influence of
the cosmological constant at high redshift. The density of absorbers
is proportional to $dN_{\rm LL}/dx$, hence we find $n_{\rm
abs}\propto(1+z)^{\beta+\frac{3}{2}}$, and thus
$\gamma=\beta+\frac{3}{2}$.  The value of $\beta$ is measured at
$z=4$. However mini-halos may be important at high redshift so that an
extrapolation of absorption properties from low to high redshift is
not justified. We therefore consider a range of values for $\beta$
which allow for a range of evolutionary properties of the absorbers as
well as their bias, and of the contribution of quasars to the ionizing
background. In the right panel of Figure~\ref{fig1} we plot the values
of the difference between $d\ln{n_{\rm abs}}/dz$ and $d\ln (dF_{\rm
col}/dt)/dz$ as a function of redshift for $\beta=0.5$, 1.5 and 2.5
($\gamma=2$, 3 and 4).

\begin{figure*}[t]
\epsscale{1.}  \plotone{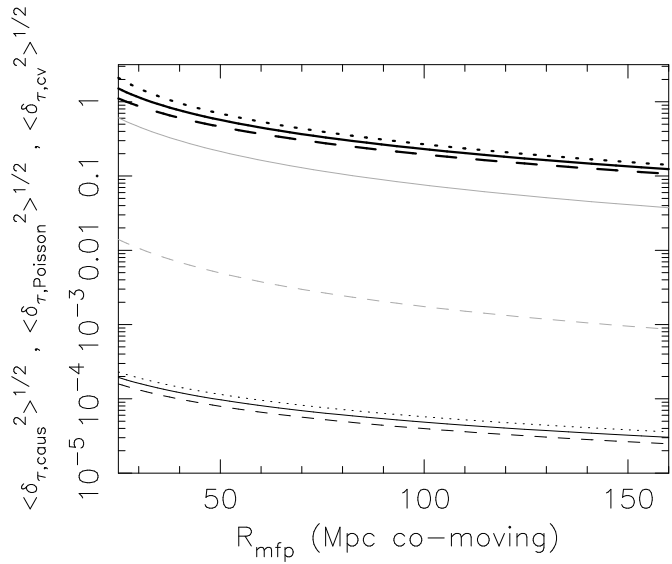}
\caption{\label{fig2} Contributions to fluctuations in $\delta_\tau$. 
The thick lines show
the fluctuations in optical depth induced through cosmic variance,
$\langle\delta_{\rm \tau,cv}^2\rangle^{1/2}$. The dashed, solid and
dotted lines correspond to $\gamma=2$, 3, and 4 respectively, and for
calculation of $F_{\rm col}$ we have assumed a minimum virial
temperature for galaxies of $2\times10^5$K. The values of
$\langle\delta_{\rm
\tau,Poisson}^2\rangle^{1/2}$ are shown by the light lines; the solid and dashed
lines correspond to $t_{\rm lt}/t_{\rm H}=0.01$ and $t_{\rm lt}/t_{\rm
H}=0.1$, combined with minimum virial temperatures for galaxies of
$2\times10^5$K and $10^4$K respectively. We also show the values of
$\langle\delta_{\rm \tau,caus}^2\rangle^{1/2}$ (dark lines with
dashed, solid and dotted curves corresponding to $\gamma=2$, 3, and 4
respectively). $\langle\delta_{\rm \tau,caus}^2\rangle^{1/2}$ is much
smaller than $\langle\delta_{\rm \tau,Poisson}^2\rangle^{1/2}$ and
$\langle\delta_{\rm \tau,cv}^2\rangle^{1/2}$. In each case we have
evaluated the fluctuations at $z=5.5$.}
\end{figure*}

In a region of density contrast $\Delta$, we can compute the typical delay due
to cosmic variance $\langle \delta z_{\rm cv}^2\rangle^\frac{1}{2}$ (see
equation~\ref{delta}). However to estimate the typical fluctuations in
the effective optical depth we must also compute the variance in
overdensity among lines of sight through the density field of length
$R_{\rm mfp}$. We calculate the power-spectrum of fluctuations in
cylinders of length $L$ and radius $R$ to second order in the wave
number $k$ (Kaiser \& Peacock~1991; Baugh~1996)
\begin{equation}
P_{\rm 1d}(k) = \frac{1}{(2\pi)^2}\frac{2\pi}{L}\int_k^\infty dy yP_{\rm b}(y)e^{-(y^2-k^2)R^2/4}, 
\end{equation}
where $P_{\rm b}(k)$ is the linear baryonic power-spectrum, which may be
approximated as $P_{\rm b}(k)=P(k)(1+k^2R_{\rm f}^2)^{-2}$ in terms of the
cold dark matter power-spectrum $P(k)$ and the filtering scale $R_{\rm f}$
for the associated reionization history (Gnedin \& Hui~1998). The variance
in $\Delta_z$ on a scale $R$ with wavenumber $k_{\rm R}$ follows from
\begin{equation}
\sigma_{R,z} = \frac{1}{2\pi}\int_0^{k_{\rm R}}dkP_{\rm 1d}(k).
\end{equation}
The radius of the cylinder $R$ is set by the size of the quasar
emission region, which is much smaller than the filtering scale. Our
results are therefore insensitive to the precise choice of $R$.  To
estimate typical fluctuations in the optical depth from cosmic
variance we assume $\Delta_z$ and $\Delta$ to be approximately
uncorrelated and combine the 1-sigma fluctuation in delay,
($\langle\delta z_{\rm cv}^2\rangle^\frac{1}{2}$), with the 1-sigma
fluctuation in density contrast, ($\Delta_z=1\pm\sigma_{\rm R,z}D$,
where $D$ is the growth factor). The typical positive (+) and negative
(-) fluctuations in optical depth due to cosmic variance may therefore
be estimated as
\begin{eqnarray}
\nonumber
\langle\delta^2_{\rm \tau,cv}\rangle^{\frac{1}{2}} &=& \left(1\pm\sigma_{\rm R,z}D\right)^2 \\
&&\hspace{-25mm}\times \left(1\pm\left(\frac{d\ln (dF_{\rm col}/dt)}{dz}-\frac{\gamma}{(1+z)}\right)\langle\delta z_{\rm cv}^2\rangle^{\frac{1}{2}}\right)^{-1}-1.
\end{eqnarray}
Figure~\ref{fig2} shows the typical positive (+) fluctuation
$\langle\delta^2_{\rm \tau,cv}\rangle^{\frac{1}{2}}$, which is monotonic in $R_{\rm mfp}$. 

\subsection{Fluctuations due to the finite light travel time}

Following the appearance of sources with luminosity density
$\Delta\times n_{\rm src}$, the ionizing intensity will grow to its
full value $J_{\nu}$ over a timescale $R_{\rm mfp}/[c(1+z)]$.  The
time when a critical $J_{\nu}$ is achieved is affected by this light
propagation delay (with ionizing sources emitting their radiation at a
retarded time). However, there may also be a scatter in the time where
the critical $J_{\nu}$ is achieved from place to place within the
region of co-moving size $R_{\rm mfp}$. This scatter will depend on
the distribution of sources.  For example, if the sources were
clustered on a scale similar to the mean-free-path, then within a
region there will be a scatter of $R_{\rm
mfp}\left[{(1+z)cdt}/{dz}\right]^{-1}$ in redshift where the critical
$J_{\nu}$ is achieved. However if the clustering is on a much smaller
scale then the scatter will be smaller.

We can estimate this scatter by considering the clustering length of
sources, $R_0$ (which is $\sim4$ co-moving Mpc for dark-matter halos with a
virial temperature of $T_{\rm vir}\sim2\times10^5$K at $z\sim5.5$). The
photons must travel a distance $R_{\rm mfp}(1+z)^{-1}$ before the critical $J_{\nu}$
is reached. The scatter in arrival times for photons coming from sources
clustered on a scale $R_0$ at a distance $R_{\rm mfp}$ may be estimated as
the crossing time of the smaller clustering length ($R_0/c$) divided by the
square-root of the number ($N_{\rm cluster}$) of clustered regions at a
distance $R_{\rm mfp}$ (where the last factor corresponds to Poisson
fluctuations in the distances of the clustered regions).  Hence the variance
in redshift where the critical $J_{\nu}$ is reached due to the light
propagation delay will therefore be
\begin{equation}
\langle\delta z_{\rm caus}^2\rangle^\frac{1}{2} = \frac{R_0}{\sqrt{N_{\rm cluster}} (1+z)} \left(\frac{cdt}{dz}\right)^{-1},
\end{equation}
where $N_{\rm cluster}$ is estimated as
\begin{equation}
N_{\rm cluster} \sim \frac{4\pi(R_{\rm mfp}/2)^2 R_0}{4\pi/3 R_0^3} = \frac{3}{4}\left(\frac{R_{\rm mfp}}{R_0}\right)^2.
\end{equation}
The corresponding redshift variance due to the light propagation delay
$\langle\delta z_{\rm caus}^2\rangle^\frac{1}{2}$ is shown on the left panel of Figure~\ref{fig1}. At a fixed density
contrast, the finite light travel time will introduce a scatter in the
value of $\delta_\tau$ of
\begin{eqnarray}
\nonumber
\langle\delta_{\rm \tau,caus}^2\rangle^\frac{1}{2} &=&\\
&&\hspace{-25mm} \left(1+\left(\frac{d\ln (dF_{\rm col}/dt)}{dz}-\frac{\gamma}{(1+z)}\right)\langle\delta z_{\rm caus}^2\rangle^\frac{1}{2}\right)^{-1}-1.
\end{eqnarray}
We find that the fluctuations in optical depth due to this effect are much
smaller than those due to cosmic variance (see Figure~\ref{fig2}).

\subsection{Fluctuations due to the finite number of ionizing sources}

The Press-Schechter formalism predicts a probability distribution (and
hence a scatter) for the number of halos within a region of a given
matter overdensity. In this paper we use the derivative of the
collapsed fraction of baryons (up to a proportionality constant) as a
proxy for the star formation rate. This quantity depends on the
overdensity $\Delta$. As a result there is a scatter in the redshift
where a given collapsed fraction is reached among representative
regions of the universe, or equivalently there is scatter among the
collapsed fractions within different regions at fixed time. In this
paper we are concerned with regions of overdensity $\Delta$ and finite
size $R_{\rm mfp}$. Therefore, in addition to the fluctuations in the
density and ionizing radiation introduced by this cosmic variance (and by
the light propagation delay), we expect a spatial variation of the
ionizing intensity due to fluctuations in the number of sources within
the finite region. We assume that these fluctuations may be
represented by Poisson noise in the number of ionizing sources within
a region of size $R_{\rm mfp}$.  We estimate the number of galaxies
using the Press-Schechter~(1974) mass function $dn_{\rm PS}/dM$ (with
the modification of Sheth \& Tormen~1999) and evaluate the number per
logarithm of mass at the minimum mass for galaxy formation ($M_{\rm
min}$).  The number of ionizing sources at a particular epoch is given
by
\begin{equation}
N_{\rm src} = \frac{t_{\rm lt}}{t_{\rm H}}\left.\frac{dn_{\rm
PS}}{d\ln M}\right|_{M_{\rm min}}\frac{4\pi}{3}\left(\frac{R_{\rm
mfp}}{2}\right)^3,
\end{equation}
where $t_{\rm lt}$ is the source lifetime and $t_{\rm H}$ is the Hubble
time at redshift $z$. In estimating the fluctuations due to variations in
the source numbers we consider two limiting cases. First, we adopt $t_{\rm
lt}/t_{\rm H}\sim0.01$ corresponding to starburst lifetimes of $\sim10^7$
years at $z\sim 5.5$ (the lifetimes of massive stars) and a minimum mass
corresponding to the Jeans mass in a photo-ionized IGM (halo virial
temperature of $\sim2\times10^5$K). This case corresponds to the smallest
source numbers and hence the largest level of Poisson fluctuations in
$J_{\nu}$. Second we consider $t_{\rm lt}/t_{\rm H}\sim0.1$ corresponding to
starburst lifetimes of $\sim10^8$ years (of order the orbital time for a
circular orbit in a star-forming disk that occupies $\sim 10\%$ of the
virial radius of its host halo). This lifetime is combined with a minimum
halo mass having a virial temperature of $\sim10^4$K, corresponding to
cooling threshold of atomic hydrogen (so that the filtering scale is
assumed not to have had time to respond to the re-heated IGM). This case
corresponds to the largest source numbers and hence to the smallest
fluctuations in $J_{\nu}$. The Poisson noise in the number of galaxies is
$\Delta N_{\rm src}=\sqrt{N_{\rm src}}$, so that the fluctuations in
the optical depth induced by Poisson fluctuations in the number of sources is
\begin{eqnarray}
\nonumber \langle\delta_{\rm \tau,Poisson}^2\rangle^\frac{1}{2} &\sim& \sqrt{\left(\frac{\partial\ln\tau}{dN_{\rm
src}}\Delta N_{\rm src}\right)^2} = \sqrt{\left(-\frac{\partial \ln J_{\nu}}{dN_{\rm
src}}\Delta N_{\rm src}\right)^2}\\ & =& \sqrt{\left(-\frac{\partial \ln N_{\rm
src}}{dN_{\rm src}}\Delta N_{\rm src}\right)^2} = \frac{1}{\sqrt{N_{\rm
src}}},
\end{eqnarray}
where we have used the relation $J_{\nu}\propto N_{\rm src}$, and
evaluated the fluctuation range of $\tau$ corresponding to the 1-sigma
level of fluctuations in the number counts of sources. The resulting
curves are plotted as a function of $R_{\rm mfp}$ in Figure~\ref{fig2}.

\subsection{Fluctuations in the optical depth}

The contributions to the redshift variance introduced by causality, by
cosmic variance and by source count fluctuations are independent of
each other. The positive (+) and negative (-) ranges of allowed
$\delta_\tau$ may therefore be estimated by adding these contributions
in quadrature,
\begin{equation}
\label{deltapm}
\langle\delta^2_\tau\rangle^\frac{1}{2} = \pm\sqrt{\langle\delta^2_{\rm \tau,cv}\rangle + \langle\delta_{\rm \tau,caus}^2\rangle + \langle\delta_{\rm \tau,Poisson}^2\rangle}.
\end{equation}

\section{Results and Discussion}

\begin{figure*}[t]
\epsscale{2.}  \plotone{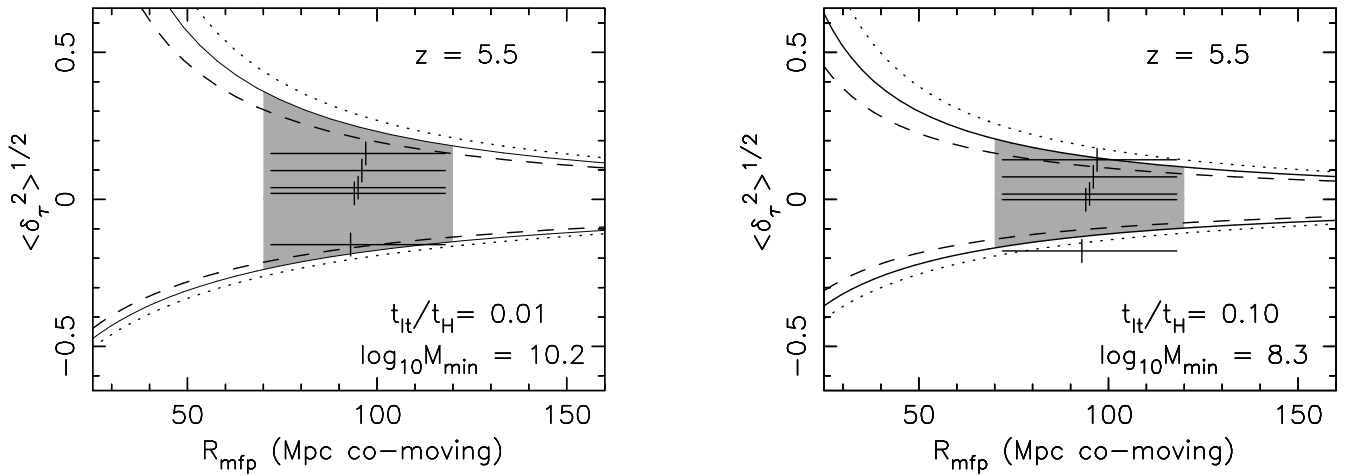}
\caption{\label{fig3} The upper and lower values of the 1-sigma range for
fluctuations in $\delta_\tau$.  The dashed, solid and dotted lines
correspond to $\gamma=2$, 3 and 4 respectively. The grey portion
corresponds to the range spanned by different values of $R_{\rm mfp}$
given $\gamma=3$. The grey region is bounded from the left by the
boundary at $70$ co-moving Mpc corresponding to the bubble size at the
end of then reionization, and from the right by the ionizing photon
mean-free-path at $z\sim4$. We show cases with $z=5.5$, and take
populations of ionizing sources having $t_{\rm lt}/t_{\rm H}=0.01$
(left) and 0.1 (right), combined with minimum galaxy masses
corresponding to virial temperatures of $\sim2\times10^5$K and
$\sim 10^4$K respectively. The values of $\delta_\tau$ at
$z\sim5.6$ measured along the lines-of-sight to five high-redshift
quasars (White et al.~2003) are over-plotted. The horizontal error-bars
on these observations indicate the possible range of $R_{\rm mfp}$.}  
\end{figure*}

The upper and lower limits that bracket the typical range of $\delta_\tau$ at
$z=5.5$ are plotted in Figure~\ref{fig3}. The dashed, solid and dotted
lines correspond to $\gamma=2$, 3 and 4 respectively. The left and
right panels show the cases of $t_{\rm lt}/t_{\rm H}=0.01$ and 0.1,
combined with minimum galaxy masses corresponding to virial
temperatures of $2\times 10^5$K and $10^4$K respectively.
Figure~\ref{fig3} shows that cosmic variance introduces fluctuations
in $\tau$ that are of order unity over a range of length scales.
As the mean free path increases the fluctuations become
smaller in line with the smaller cosmic variance on large scales.  The
fluctuations in $J_{\nu}$ introduced by fluctuations in the number of
sources provides a small contribution. This is in contrast to the situation
for double reionization of helium at lower redshifts where the ionizing
sources (quasars) reside in rarer more massive systems (Bolton et
al. 2005).

The finding that
cosmic variance should result in significant fluctuations in $\tau$ is
qualitatively insensitive to variations in the value of $\gamma$
(which parameterizes unknowns regarding the absorber properties) over
a wide range. This insensitivity arises partly because the collapsed
fraction evolves much more rapidly than a power-law with redshift
(because the minimum galaxy mass lies on the exponential tail of the
Gaussian density distribution), and partly because the fluctuations
are dominated by the $\Delta_z^2$ term in the evaluation of $\tau$.

We may take the upper limit for $R_{\rm mfp}$ after reionization to
be $120$ co-moving Mpc, which is the value measured at $z\sim4$
(Miralda-Escude et al.~2003), while the lower limit is set by the size
of bubbles at the end of the overlap epoch ($\sim70$ co-moving Mpc at
$z\sim 6$ and larger if reionization completed earlier; Wyithe \&
Loeb~2004b).  The shaded region in Figure~\ref{fig3} delineates the
range for different values of $R_{\rm mfp}$ assuming $\gamma=3$. We
find our main conclusion, that cosmic variance introduces large
fluctuations in the transmission of the IGM at $z\sim5.5$, holds over
this full allowed range of Ly-limit mean-free-paths.
The reionization redshift itself is also inhomogeneous, as evidenced
by the transmission of flux in the Ly$\alpha$ troughs of some but not
all of the $z>6$ quasars (White et al.~2003; Oh \&
Furlanetto~2004). Our model becomes applicable at the redshift where
most lines of sight probe the post-overlap IGM. Finally, we note that
our results do not depend on the reionization redshift having occurred
near $z\sim6$ but apply to earlier reionization epochs as well.

\subsection{Scatter in the transmitted intensity}

The fluctuation in the relative intensity ($I\propto e^{-\tau}$) transmitted
through the IGM ($\delta_{I}=(I-I_{\rm av})/I_{\rm av}$) is related to the
fluctuation in $\tau$ through
\begin{eqnarray}
\label{delI}
\nonumber
\delta_I&=&\frac{I-I_{\rm av}}{I_{\rm av}}=\frac{e^{-(\tau_{\rm av}+\tau_{\rm
av}\delta_\tau)}-e^{-\tau_{\rm av}}}{e^{-\tau_{\rm av}}}\\ 
&=&e^{-\tau_{\rm av}\delta_\tau}-1.
\end{eqnarray}
Note that in regions where the optical depth is reduced ($\delta_\tau<0$)
the exponential is positive and the value of $\delta_I$ can become very
large. For values of optical depth larger than unity the exponential acts
as a non-linear amplifier of any inhomogeneities in the IGM which skews the
distribution of transmitted intensity. Cosmic variance therefore leads to a
highly asymmetric distribution of transmitted flux, even if the
distribution of $\delta_\tau$ is a symmetric Gaussian.  
For large values of $\tau\sim3$--5 immediately following reionization, the
variations in the transmission span an order of magnitude. At later times
when the optical depth becomes small, equation~(\ref{delI}) becomes
\begin{equation}
\delta_I\sim-\tau_{\rm av}\delta_\tau.
\end{equation}
Thus at small values of optical depth the distribution of transmitted
intensity becomes symmetric.

\subsection{Comparison with observations}

The results presented in Figure~\ref{fig3} may be compared with existing
observations of high redshift quasars. Becker et al.~(2001) and White et
al.~(2003) calculated the effective Ly$\alpha$ optical depth in bins of
width $\Delta z=0.2$ along the lines of sight to five high redshift
quasars. The resulting values of $\delta_\tau$ for these five lines of
sight in the redshift bin $5.5\leq z\leq5.7$ are plotted over the grey
region in Figure~\ref{fig3} for comparison. At $z\sim5.6$ White et
al.~(2003) find $\tau_{\rm av}\sim2.5$ and fractional fluctuations in
optical depth with values up to $(\tau-\tau_{\rm av})/\tau_{\rm
av}\sim0.25$. Since there are only 5 lines of sight we would expect this
observed range to be roughly equivalent to the 1-sigma fluctuations plotted
in figure~\ref{fig3}, for which we find $\langle\delta^2_\tau\rangle^\frac{1}{2}\sim0.1-0.4$ depending
on the value of the mean-free-path and the nature of the ionizing
sources. Inspection of Figure~\ref{fig3} therefore suggests that the model
is in good agreement with with current data.  Djorgovski~(2004) has also
presented the fluctuations in optical depth along five lines of sight
within Gaussian redshift windows of variance $\sigma_z=0.05$. This smaller
redshift window yields fractional fluctuations in optical depth with values
up to $(\tau-\tau_{\rm av})/\tau_{\rm av}\sim0.5$. At $z\sim5.5$, the
redshift bin widths of $\Delta z=0.1-0.2$ correspond to co-moving
line-of-sight distances $R_{\Delta z}=(1+z)cdt/dz \Delta
z\sim50-100$Mpc. Thus the observations sample the IGM over a length-scale
comparable to $R_{\rm mfp}$, and so are fortuitously matched to our
calculation. Of course at redshifts where $R_{\rm mfp}<50-100$Mpc, our
estimate of $\langle\delta^2_\tau\rangle^\frac{1}{2}$ may be compared with observation by averaging
the predicted fluctuations over a number $\sim [(50-100\mbox{Mpc})/R_{\rm
mfp}$] of regions.

\subsection{Relative absorption of Ly$\alpha$ and Ly$\beta$ photons}
 
The relative level of absorption for Ly$\alpha$ and Ly$\beta$ (or
Ly$\gamma$ etc) photons is dictated by the ratio of the product of
oscillator strengths and resonance wavelengths for the two
transitions. Assuming a homogeneous IGM this results in a Ly$\alpha$
optical depth that is 6.25 times the Ly$\beta$ optical depth
(Songaila~2004). However the ratio is decreased in an inhomogeneous
IGM because on small scales transmission is dominated by low-density
regions, while the transition with the lower cross-section for absorption is
able to sample a larger range of densities. Oh \& Furlanetto~(2004)
find that values of 2-2.5 are more realistic for the ratio of Ly$\alpha$ to
Ly$\beta$ optical depth.

The constant of proportionality relating $\tau$ and $J_{\nu}$ will
therefore be dependent on the oscillator strength and resonance
frequencies. This constant cancels in calculation of $\delta_\tau$,
and so the range of fractional variation in optical depth will be the
same for absorption in Ly$\alpha$ and Ly$\beta$. On the other hand the
range of optical depth values is given by $\tau_{\rm
av}\langle\delta^2_\tau\rangle^\frac{1}{2}$. As a result, the absolute range of optical depths
observed for Ly$\alpha$ absorption will be larger than the range
observed for Ly$\beta$ absorption by a factor of $\sim
2-6$. Similarly, while the relative fluctuations in the optical depth
are the same for all transitions, we find that the relative
fluctuations in observed intensity $\delta_I$ will be much larger for
Ly$\alpha$ than Ly$\beta$ photons due to the non-linearity of the
exponential relation between optical depth and transmission. In
particular, if the Ly$\alpha$ optical depth is smaller than unity,
then $\delta_I\sim-\tau_{\rm av}\delta_\tau$ and the relative
fluctuations in intensity may be calculated directly from the ratios
of the optical depths among the different transitions.  In cases where
the Ly$\alpha$ optical depth is larger than unity, the variation in
$\delta_I$ among the different transitions is larger due to the
presence of the exponential in equation~(\ref{delI}).

\vspace{5mm}
\section{Summary}

In this paper we have calculated the magnitude of the large scale
fluctuations in the optical depth to absorption of Ly-series photons that
are introduced through variance in the cosmic density field. We find
that cosmic variance introduces fractional fluctuations in the optical
depth that are of order unity for mean-free paths of ionizing photons
$R_{\rm mfp}\la150$ co-moving Mpc. This finding explains the variation seen
in the Ly$\alpha$ optical depth along different lines of sight through the
IGM at $z<6$ (Becker et al.~2001; White et al.~2003).

Our model was not constructed to predict the value of optical depth or
to associate a value of the optical depth with the
mean-free-path. Rather, it predicts the trend that the relative
fluctuations in optical depth will drop as the ionizing photon
mean-free-path grows. The mean-free-path increases with cosmic time,
and so the range of optical depth will therefore decrease towards low
redshift. In addition, while our model predicts a distribution of
optical depths that is roughly symmetric about the average, the
exponential dependence of transmission on optical depth implies a
distribution of transmission that will be highly asymmetric. This
asymmetry will become smaller as the average value of optical depth
declines at late times. Finally, our model predicts that the absolute
fluctuations in the Ly$\alpha$ optical depth should be greater than in
the Ly$\beta$ optical depth.  This translates to relative fluctuations
in the transmission through the IGM that are much larger for
Ly$\alpha$ than for Ly$\beta$.

As larger numbers of high redshift quasars are discovered through surveys
like the {\it Sloan Digital Sky Survey}, the statistics of absorption
averaged over large scales will be measured along many lines of sight. The
results of this paper suggest that the effects of cosmic variance on the
post-reionization epoch will be directly observed in these studies.

\acknowledgements 

The authors would like to thank Xiaohui Fan for pointing out an error
in an early version of this manuscript. JSBW acknowledges the support
of the Australian research council. This work was supported in part by
NASA grants NAG 5-13292 and NNG05GH54G for AL.

\end{document}